\documentclass[showpacs]{revtex4}
\usepackage{epsfig}
\def\beq{\begin{equation}}
\def\enq{\end{equation}}
\def\beqa{\begin{eqnarray}}
\def\enqa{\end{eqnarray}}

\def\GeV{\nobreak\,\mbox{GeV}}

\def\qq{\lag\bar{q}q\rag}

\def\ss{\lag\bar{s}s\rag}

\def\Gd{\lag g^2G^2\rag}
\def\G3{\lag g^3G^3\rag}

\def\rh{\rho}
\def\si{\sigma}

\def\al{\alpha}
\def\be{\beta}
\def\alma{\alpha_{max}}
\def\almi{\alpha_{min}}
\def\bemi{\beta_{min}}
\def\mme{m_{D_sD^*}}
\def\lb{\label}
\def\nn{\nonumber}

\newcommand{\rag}{\rangle}
\newcommand{\lag}{\langle}

\begin{document}

\title{\sc $D_sD^*$ molecule as an axial meson}
\author{Su Houng Lee}
\email{suhoung@phya.yonsei.ac.kr}
\affiliation{Institute of Physics and Applied Physics, Yonsei University,
Seoul 120-749, Korea}
\author{Marina Nielsen}
\email{mnielsen@if.usp.br}
\affiliation{Instituto de F\'{\i}sica, Universidade de S\~{a}o Paulo,
C.P. 66318, 05389-970 S\~{a}o Paulo, SP, Brazil}
\author{Ulrich Wiedner}
\email{ulrich.wiedner@ruhr-uni-bochum.de}
\affiliation{Bochum University, Inst. Exp. Phys. I, Gbd,NB 2/131, D-44789
Bochum, Germany}

\begin{abstract}
We use QCD  sum rules to study the possible existence of a $D_s\bar{D}^*+
D_s^*\bar{D}$ molecule with the quantum number $J^P=1^+$.
We consider the contributions of condensates up to dimension eight and
work at leading order in $\alpha_s$. We obtain $m_{D_sD^*}=(3.96\pm0.10)~\GeV$
around 100 MeV above the mass of the meson $X(3872)$.  The proposed state is
a natural generalized state to the strangeness sector of the $X(3872)$, which
was also found to be consistent with a multiquark state from a previous QCD
sum rule analysis.
\end{abstract}

\maketitle


The recent appearance of new resonances observed
by BaBar, BELLE, CLEO and FOCUS collaborations, and partly confirmed by
the Fermilab experiments CDF and D0, sheds new light on the spectroscopy
of charmonium states. Among these new mesons, some have their masses
very close to the meson-meson threshold like the $X(3872)$ \cite{belle1}
and the $Z^+(4430)$ \cite{belle2}. Of special importance is the
appearance of the $Z^+(4430)$, which decays into $\psi^\prime$ and $\pi^+$
and, therefore, can not be described as ordinary $c\bar{c}$ meson. Its
nature is completely open, but an intriguing possibility is the interpretation
as tetraquark state or molecular state \cite{swan,meng,nos}.
If these mesons are really molecular
states, then many other molecules should exist.   A systematic study of these
molecular states and their experimental observation would confirm its
structure and provide a new testing ground for QCD within multiquark
configuration.    In this context a natural extension would be to probe the
strangeness sector.
In particular, in analogy with the meson $X(3872)$, a $D_sD^*$ molecule with
$J^P=1^+$ would decay into $J/\psi K^*\to J/\psi K\pi$ and, therefore, could
be easily reconstructed. Here we use the QCD sum rules \cite{svz,rry,SNB}
to predict the mass of such a molecular state.

This calculation is of particular importance for new upcoming experiments
which can investigate with much higher precision the charmonium energy regime,
like the PANDA experiment at the antiproton-proton facility at FAIR, or a
possible Super-B factory experiment. Especially PANDA can do a careful scan of
the various thresholds being present, in addition to precisely going through
the exact form of the resonance curve.

In a previous calculation, the QCDSR approach was used to study
the $X(3872)$ considered as a diquark-antidiquark state \cite{x3872},
and the $Z^+(4430)$ meson, considered as a $D^*D_1$ molecular state \cite{nos}.
In both cases a very good agreement with the experimental mass was obtained.


Considering a $D^*D_s$ molecule with $J^P=1^+$, a possible
current describing such state is given by:
\beq
j_\mu={i\over\sqrt{2}}\left[(\bar{s}_a\gamma_5 c_a)(\bar{c}_b\gamma_\mu
d_b)-(\bar{s}_a\gamma_\mu c_a)(\bar{c}_b\gamma_5 d_b)\right]\;,
\label{field}
\enq
where $a$ and $b$ are color indices. We have considered the anti-symmetrical
state $D^{*-}D_s^+-D^-{D_s}^{*+}$ to keep a closer relation with the
$X(3872)$ meson. Considering the $X(3872)$ as a $D^*D$ molecule, the
combination
$D^{*0}\bar{D}^0-D^0\bar{D}^{*0}$ has $J^{PC}=1^{++}$ as the $X(3872)$ meson.
Of course the symmetrical combination: $D^{*-}D_s^++D^-{D_s}^{*+}$ would
provide exactly the same mass, within our sum rule approach.

The sum rule is constructed from the two-point correlation function:
\beq
\Pi_{\mu\nu}(q)=i\int d^4x ~e^{iq.x}\lag 0
|T[j_\mu(x)j_\nu^\dagger(0)]|0\rag=-\Pi_1(q^2)(g_{\mu\nu}-{q_\mu q_\nu
\over q^2})+\Pi_0(q^2){q_\mu q_\nu\over q^2}.
\lb{2po}
\enq
Since the axial vector current is not conserved, the two functions,
$\Pi_1$ and $\Pi_0$, appearing in Eq.~(\ref{2po}) are independent. They
have respectively the quantum numbers of the spin 1 and 0 mesons. Therefore,
we choose to work with the Lorentz structure $g_{\mu\nu}$, since it
gets contributions only from the $1^{+}$ state.

On the OPE side, we work at leading order in $\alpha_s$ in the operators and
consider the contributions from condensates up to dimension eight.
The correlation function, $\Pi_1$,  in the OPE side can be written as a
dispersion relation:
\beq
\Pi_1^{OPE}(q^2)=\int_{4m_c^2}^\infty ds {\rho^{OPE}(s)\over s-q^2}
+\Pi_1^{m_s\qq^2}(q^2)+\Pi_1^{mix\qq}(q^2)\;,
\lb{ope}
\enq
where $\rho^{OPE}(s)$ is given by the imaginary part of the
correlation function: $\pi \rho^{OPE}(s)=\mbox{Im}[\Pi_1^{OPE}(s)]$.
We get:
\beq
\rho^{OPE}(s)=\rho^{pert}(s)+\rh^{m_s}(s)+\rh^{\qq}(s)+\rh^{m_s\qq}(s)
+\rh^{\lag G^2\rag}(s)+\rh^{mix}(s)+\rh^{\qq^2}(s)+\rh^{m_s.mix}(s)+
\rh^{m_s\qq^2}(s)\;,
\lb{rhoeq}
\enq
with
\beqa\label{eq:pert}
&&\rho^{pert}(s)={3\over 2^{12} \pi^6}\int\limits_{\almi}^{\alma}
{d\al\over\alpha^3}
\int\limits_{\bemi}^{1-\al}{d\be\over\be^3}(1-\al-\be)(1+\al+\be)
\left[(\al+\be)m_c^2-\al\be s\right]^4,
\nn\\
&&\rho^{ms}(s)=-{3m_sm_c \over 2^{10} \pi^6} \int\limits_{\almi}^{\alma}
{d\al\over\al^3} {d\be\over\be^2}(1-\al-\be)(3+\al+\be)\left[(\al+\be)m_c^2
-\al\be s\right]^3,
\nn\\
&&\rho^{\qq}(s)=-{3m_c\over 2^{8}\pi^4}(\qq+\ss)\int\limits_{\almi}^{\alma}
{d\al\over\al^2}
\int\limits_{\bemi}^{1-\al}{d\be\over\be}(1+\al+\be)\left[(\al+\be)m_c^2-
\al\be s\right]^2,
\nn\\
&&\rho^{m_s\qq}(s)={3m_s\over 2^{8}\pi^4}\int\limits_{\almi}^{\alma}
{d\al\over\al}\Bigg[{\left((\al+\be)m_c^2-\al\be s\right)^2\over1-\al}\ss+
\int\limits_{\bemi}^{1-\al}{d\be\over\be}\left((\al+\be)m_c^2-
\al\be s\right)\left(4m_c^2\qq\right.
\nn\\
&&\left.-\left((\al+\be)m_c^2-\al\be s\right)\ss\right)
\Bigg],
\nn\\
&&\rho^{\lag G^2\rag}(s)={\Gd\over2^{11}\pi^6}\int\limits_{\almi}^{\alma}
d\al\!\!\int\limits_{\bemi}^{1-\al}{d\be\over\be^2}\left[(\al+\be)m_c^2-\al\be
s\right]\left[{m_c^2(1-(\al+\be)^2)\over\be}-
{(1-2\al-2\be)\over\al}\left[(\al+\be)m_c^2-\al\be s\right]
\right],
\nn\\
&&\rho^{mix}(s)={3m_cm_0^2\over 2^{9}\pi^4}\left(\qq+\ss\right)
\int\limits_{\almi}^{\alma}d\al
\bigg[-{2\over\al}(m_c^2-\al(1-\al)s)
+\int\limits_{\bemi}^{1-\al}d\be\left[(\al+\be)m_c^2-\al\be
s\right]\left({1\over
\al}+{2(\al+\be)\over\be^2}\right)\bigg],
\nn\\
&&\rho^{\qq^2}(s)={m_c^2\qq\ss\over 2^4\pi^2}\sqrt{1-4m_c^2/s},
\nn\\
&&\rho^{m_s.mix}(s)={m_sm_0^2\over 2^{8}\pi^4}\Bigg\{
\sqrt{1-4m_c^2/s}\left(6m_c^2\qq+{m_c^2+2s\over 3}\ss\right)-
\int\limits_{\almi}^{\alma}d\al\Bigg[3m_c^2\qq\left({1\over\al}-{m_c^2\over\al
s-m_c^2}\right)+
\nn\\
&&{m_c^2\over1-\al}\ss-2\ss
\int\limits_{\bemi}^{1-\al}{d\be\over\be}\left[(\al+\be)m_c^2-\al\be
s\right]\Bigg]\Bigg\},
\nn\\
&&\rho^{m_s\qq^2}(s)={3m_sm_c\qq\ss\over 2^7\pi^2}\sqrt{1-4m_c^2/s},
\nn\\
&&\Pi_1^{m_s\qq^2}(q^2)=-{m_sm_c^3\qq\ss\over 2^5\pi^2}\int_0^1
d\al{1-\al\over( m_c^2-\al(1-\al)q^2},
\nn\\
&&\Pi_1^{mix\qq}(q^2)=-{m_c^2m_0^2\qq\ss\over 2^5\pi^2}\int_0^1
d\al{\al(1-\al)\over m_c^2-\al(1-\al)q^2}\left[1+{m_c^2
\over m_c^2-\al(1-\al)q^2}-{1\over1-\al}\right].
\label{dim8}
\enqa
where the integration limits are given by $\almi=({1-\sqrt{1-
4m_c^2/s})/2}$, $\alma=({1+\sqrt{1-4m_c^2/s})/2}$, $\bemi={\al
m_c^2/( s\al-m_c^2)}$, and we have used $\lag\bar{q}g\si.Gq\rag=m_0^2\qq$,
$\lag\bar{s}G\si.GS\rag=m_0^2\ss$.

It is very interesting to notice that the current in Eq.~(\ref{field}) has
a OPE behavior very similar to the scalar-diquark axial-antidiquark current
used for the $X(3872)$ meson in ref.~\cite{x3872}.

In the phenomenological side, we write a dispersion relation to the
correlation function in Eq.~(\ref{2po}):
\beq
\Pi_1^{phen}(q^2)=\int ds\, {\rho^{phen}(s)\over s-q^2}\,+\,\cdots\,,
\label{phen}
\enq
where $\rho^{phen}(s)$ is the spectral density and the dots represent
subtraction terms. The spectral density is described, as usual, as a single
sharp
pole representing the lowest resonance plus a smooth continuum representing
higher mass states:
\beqa
\rho^{phen}(s)&=&\lambda^2\delta(s-\mme^2) +\rho^{cont}(s)\,,
\label{den}
\enqa
where $\lambda$ gives the coupling of the current to the meson $D_sD^*$:
\beq\label{eq: decay}
\lag 0 |
j|D_sD^*\rag =\lambda.
\enq

For simplicity, it is
assumed that the continuum contribution to the spectral density,
$\rho^{cont}(s)$ in Eq.~(\ref{den}), vanishes bellow a certain continuum
threshold $s_0$. Above this threshold, it is assumed to be given by
the result obtained with the OPE. Therefore, one uses the ansatz \cite{io1}
\beq
\rho^{cont}(s)=\rho^{OPE}(s)\Theta(s-s_0)\;,
\enq

 After making a Borel transform to both sides of the sum rule, and
transferring the continuum contribution to the OPE side, the sum rules
for the pseudoscalar meson $Z^+$, up to dimension-eight condensates, can
be written as:
\beq \lambda^2e^{-\mme^2/M^2}=\int_{4m_c^2}^{s_0}ds~
e^{-s/M^2}~\rho^{OPE}(s)\; +\Pi^{m_s\qq^2}(M^2)\; +\Pi^{mix\qq}(M^2)\;, \lb{sr}
\enq

To extract the mass $\mme$ we take the derivative of Eq.~(\ref{sr})
with respect to $1/M^2$, and divide the result by Eq.~(\ref{sr}).

\begin{figure}[h]
\centerline{\epsfig{figure=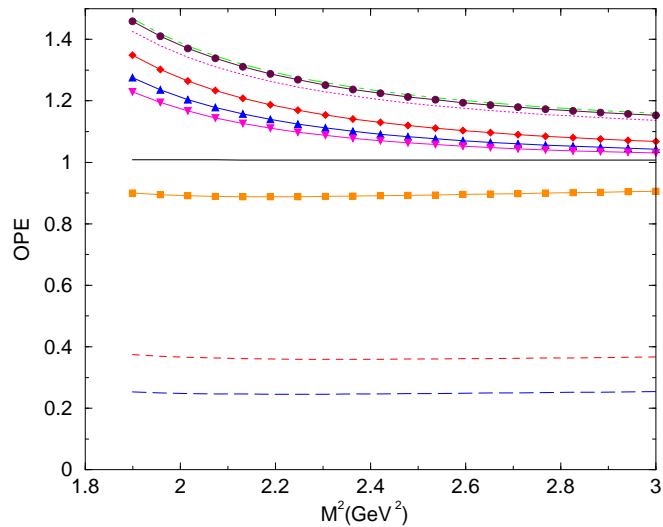,height=70mm}}
\caption{The OPE convergence in the region $1.9 \leq M^2 \leq
3.0~\GeV^2$ for $\sqrt{s_0} = 4.5$ GeV. We start with the relative
perturbative contribution (long-dashed line),
and each other line represents the relative contribution after adding of one 
extra condensate in the expansion: +$m_s$ (dashed line), + $\qq$ (dotted
line), + $m_s\qq$ (dot-dashed line), + $\langle g^2G^2\rangle$ (line with
circles), + $m_0^2\qq$ (line with squares), + $\qq^2$ (line with diamonds),
+ $m_sm_0^2\qq$ (line with triangles up), + $m_s\qq^2$ (line with triangles
down), + $m_0^2\qq^2$ (solid line).}
\label{figconvtcc}
\end{figure}

The values used for the quark
masses and condensates are \cite{SNB,narpdg}:
$m_c(m_c)=(1.23\pm 0.05)\,\GeV $,
$\lag\bar{q}q\rag=\,-(0.23\pm0.03)^3\,\GeV^3$, $\ss=0.8\qq$,
$m_0^2=0.8\,\GeV^2$, $\lag g^2G^2\rag=0.88~\GeV^4$.

We evaluate the sum rules in the Borel range $1.9 \leq M^2 \leq 3.5\GeV^2$,
and in the $s_0$ range $4.4\leq \sqrt{s_0} \leq4.6$ GeV.

From Fig.~\ref{figconvtcc} we see that we obtain a reasonable OPE
convergence for $M^2\geq 1.9$ GeV$^2$. Therefore, we  fix the lower
value of $M^2$ in the sum rule window as $M^2_{min} = 1.9$ GeV$^2$.
We notice that the OPE convergence in this case is similar to the OPE
convergence for the $X(3872)$ meson \cite{x3872}, and not so good as the OPE
convergence for the $Z^+(4430)$ meson \cite{nos}.

\begin{figure}[h]
\centerline{\epsfig{figure=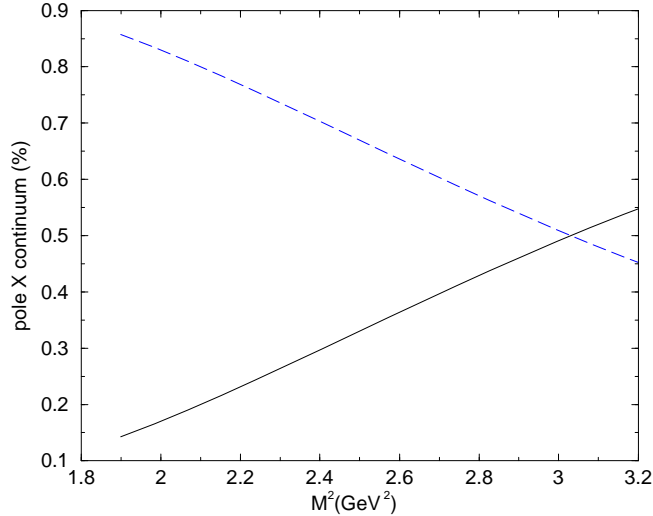,height=70mm}}
\caption{The dashed line shows the relative pole contribution (the
pole contribution divided by the total, pole plus continuum,
contribution) and the solid line shows the relative continuum
contribution for $\sqrt{s_0}=4.5~\GeV$.}
\label{figpvc}
\end{figure}

To get an upper limit constraint for $M^2$ we impose that
the QCD continuum contribution should be smaller than the pole contribution.
The comparison between pole and
continuum contributions for $\sqrt{s_0} = 4.5$ GeV is shown in
Fig.~\ref{figpvc}. From this figure we see that the pole contribution
is bigger than the continuum for $M^2\leq3.0~\GeV^2$.
The maximum value of $M^2$ for which this constraint is satisfied
depends on the value of $s_0$. The same analysis for the other values of the
continuum threshold gives $M^2 \leq 2.8$  GeV$^2$ for $\sqrt{s_0} = 4.4~\GeV$
and $M^2 \leq 3.2$  GeV$^2$ for $\sqrt{s_0} = 4.6~\GeV$.
In our numerical analysis, we shall then consider the range of $M^2$ values
from 1.9 $\GeV^2$ until the one allowed by the pole dominance criteria given
above.

\begin{figure}[h]
\centerline{\epsfig{figure=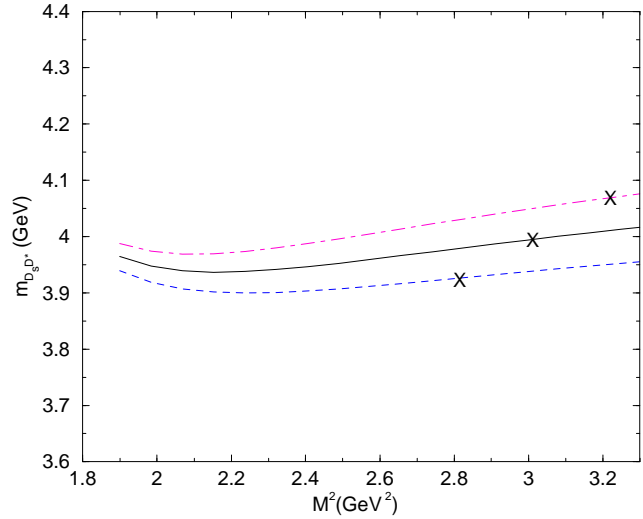,height=70mm}}
\caption{The $D_sD^*$ meson mass as a function of the sum rule parameter
($M^2$) for $\sqrt{s_0} =4.4$ GeV (dashed line), $\sqrt{s_0} =4.5$ GeV (solid
line) and $\sqrt{s_0} =4.6$ GeV (dot-dashed line). The crosses
indicate  the upper limit in the Borel region allowed
by the dominance of the QCD pole contribution.}
\label{figmx}
\end{figure}
In Fig.~\ref{figmx}, we show the $D_sD^*$ meson mass, for different values
of $\sqrt{s_0}$, in the relevant sum rule window, with the upper validity
limits indicated.  From this figure we see that
the results are very stable as a function of $M^2$.

Using the Borel window, for each value of $s_0$, to evaluate the mass of the
$D_sD^*$ meson and then varying the value of the continuum threshold in the
range $4.4\leq \sqrt{s_0} \leq4.6$
GeV, we get
\beq
\mme = (3.97\pm0.08)~\GeV,
\label{zmass}
\enq
around 100 MeV bigger than the mass of the $X(3872)$ meson.
The Borel curve for the mass is quite stable and has a minimum within the
relevant Borel window.  Such stable Borel curve strongly suggests that there
is indeed a very well defined ground state.

To check the dependence of our results with the value of the
charm quark mass, we fix $\sqrt{s_0}=4.5~\GeV$ and vary the charm quark mass
in the range $m_c=(1.23\pm0.05)~\GeV$. Using $1.9\leq M^2\leq 3.0~\GeV^2$
we get: $\mme = (3.96\pm0.10)~\GeV$, in agreement
with the result in Eq.~(\ref{zmass}). Therefore, we conclude that the most
important sources of uncertainty in our calculation are the values
of the continuum threshod and the charm quark mass.

In conclucion, we have presented a QCDSR analysis of the two-point
function for a possible  $D_s\bar{D}^*+ D_s^*\bar{D}$ molecular state with
$J^P=1^+$.
This state would decay into $J/\psi K^*\to J/\psi K\pi$ and, therefore, could
be easily reconstructed.  Our finding strongly suggests the possibility of
the existence of such molecular resonance, whose structure is similar to the
$X(3872)$, with a mass about 100 MeV above that of $X(3872)$.

The mass of our proposed state is also close to the newly discovered 
$X(3940)$ \cite{X3940} and $Z(3930)$ \cite{Z3930}.   While our proposed state has 
stangeness, the two states do not.
Moreover, while $X(3940)$ was found in the decay of $D^*\bar{D}$ and is a 
candidate for $\eta_c^{''}(3 ^1S_0)$, $Z(3930)$ was found in $D \bar{D}$ and is
consistent with $\chi_{c2}^\prime$.  Therefore,  a comparative study with these 
particles can give us clues on the structures of $X(3872)$ and our proposed 
state.

As a final remark, one can notice that if we take $m_s=0$, one obtains the
sum sule for the $X(3872)$ meson, considered as a molecular state
$D^0\bar{D}^{*0}- D^{*0}\bar{D}^0$. The results are very similar to the ones
obtained in ref.~\cite{x3872}, but more stable as a function of the Borel mass.
Using $\sqrt{s_0}=(4.4\pm0.1)~\GeV$ one obtains $m_X=(3.88\pm0.06)~\GeV$,
in an excelent agreement with the experimental value.

\section*{Acknowledgements}

We would like to thank APCTP for sponsoring the workshop on 'Hadron Physics
at RHIC'.  The discussions during the workshop have lead the authors to
collaborate on this subject.
This work has been partly supported by FAPESP and CNPq-Brazil,
by the Korea Research Foundation KRF-2006-C00011 and by the German BMBF grant 
06BO108I.


\end{document}